# TriQuest：An AI Copilot-Powered Platform for Interdisciplinary Curriculum Design

Huazhen Wang[1], Huimin Yang[1], Hainbin Lin[1], Yan Dong[2], Lili Chen[2], Liangliang Xia[2], Wenwen Xu[2]

(1. College of Computer Science and Technology, Huaqiao University, Xiamen, Fujian 361021)

(2. Faculty of Education, Beijing Normal University, Beijing 100875)

## Abstract

As curriculum reforms deepen, interdisciplinary teaching has emerged as a central theme in educational advancement. However, educators commonly encounter significant challenges, including difficulties in integrating knowledge across disciplines and the time-intensive nature of developing high-quality lesson plans. Current teaching aids and general AI models are often inadequate for the intricate requirements of interdisciplinary teaching design, particularly concerning deep domain knowledge, pedagogical congruence, and alignment with curriculum standards. To address these challenges, we introduce TriQuest, a platform meticulously designed to empower educators in efficiently and effectively generating high-quality interdisciplinary lesson plans. Operating on an AI-copilot model, the platform leverages human-computer interaction technologies, including large language models and knowledge graph technology, via an intuitive Graphical User Interface (GUI). It intelligently integrates knowledge from various disciplines, utilizes a built-in prompt template library to construct lesson plan generation templates, and then employs large language models to produce high-quality interdisciplinary lesson plans. Throughout this process, a deeply integrated human-computer collaborative review and feedback mechanism critically ensures the quality, innovativeness, and pedagogical agency of the created lesson plans. Our user research, involving 43 frontline teachers, demonstrates that the TriQuest platform significantly enhances teachers' efficiency in interdisciplinary curriculum design, improves the interdisciplinary integration and innovativeness of lesson plans, and concurrently lowers design barriers and cognitive load for educators. This work establishes a novel paradigm and provides empirical evidence for the profound empowerment of teacher professional development through intelligent technologies.

## 1 Introduction

In recent years, global educational reforms have consistently moved towards a common trend: dismantling disciplinary boundaries to foster interdisciplinary learning. Specifically, China's "Compulsory Education Curriculum Plan and Curriculum Standards (2022 Edition)" mandates that at least 10% of class hours in each subject be dedicated to interdisciplinary thematic learning, marking a shift from theoretical advocacy to a stringent requirement for interdisciplinary instruction（Ministry of Education，2022）. The fundamental objective of this approach is to cultivate students' ability to address complex real-world problems by integrating multidisciplinary knowledge, perspectives, and methods, thereby elevating their core competencies (Dong et al., 2023; Gu et al., 2023).

Nevertheless, this ambitious educational vision confronts significant practical challenges. China's teacher training system has historically emphasized single-subject education, leaving many frontline educators without the requisite interdisciplinary academic bac kground or teaching experience. This often leads to practical challenges, characterized by "interdisciplinarity for its own sake" or fragmented integration (Yang et al., 2024). When tasked with designing interdisciplinary lesson plans, teachers frequently struggle with selecting appropriate integrative themes, seamlessly combining knowledge from disparate subjects, and developing coherent, in-depth inquiry activities. The manual creation of high-quality interdisciplinary lesson plans is not only immensely time-consuming but also places exceptionally high demands on teachers' breadth of knowledge and capacity for pedagogical innovation. Consequently, effectively empowering teachers, thereby lowering the entry barrier for interdisciplinary teaching design and simultaneously enhancing its efficiency and quality, has become a critical bottleneck in the successful implementation of the new curriculum reform (Li et al., 2024).

To address this challenge, both academia and industry have explored various solutions. Traditional methods, such as expert lectures and collaborative teaching research, offer some insights but are limited in scope, costly, and often fail to provide personalized, process-oriented support. More recently, Generative Artificial Intelligence (GAI), particularly Large Language Models (LLMs), has showcased robust content generation capabilities, opening new avenues for the education sector (Yu et al., 2023). Some educators have begun experimenting with general-purpose AI tools (e.g., ChatGPT, Wenxin Yiyan) to aid in curriculum design. However, these general models exhibit notable limitations: primarily, their training on broad internet data means they often lack a deep understanding of specific national curriculum standards, textbook editions, and pedagogical theories. Consequently, the content they generate can be "off-topic" and inconsistent with teaching guidelines. Secondly, their typical "one-stop" text output often falls short in supporting the complex, iterative creative process inherent in instructional design, potentially leading to teachers' passive acceptance and diminishing their pedagogical agency. While existing teaching and research platforms contain numerous cases, these are predominantly static resources, resulting in inefficient retrieval and reuse.

In summary, there is a pressing demand for a comprehensive, end-to-end solution explicitly tailored for interdisciplinary pedagogy. To address this critical need, we have conceptualized and developed the TriQuest platform. This platform represents a unified and extensible framework, specifically engineered to empower educators in the efficient and high-quality design of interdisciplinary lesson plans through profound human-computer collaboration. This approach seeks to optimize the instructional design process, ensuring both effectiveness and pedagogical excellence.

Our primary contributions are summarized as follows:

- We introduce an innovative, end-to-end framework, TriQuest, meticulously crafted for the development of interdisciplinary lesson plans. This framework systematically integrates domain-specific knowledge graphs, guided human-computer interaction, and multi-module generation technologies, thereby effectively mitigating the deficiencies of existing tools concerning professional specificity, comprehensive process support, and teacher agency.
- We have developed a visualized platform leveraging an AI-copilot paradigm, which substantially reduces the entry barrier for educators without technical backgrounds to harness advanced AI technologies. This platform, through its guided workflows, externalizes and structures intricate instructional design thought processes, thereby fostering profound professional reflection and continuous growth among teachers.
- Through a rigorous user study involving 43 frontline educators, we substantiated the platform's efficacy across both qualitative and quantitative measures. The empirical findings unequivocally demonstrate that the platform yields significant improvements in curriculum design efficiency, elevates the overall quality of lesson plans, and markedly stimulates teachers' propensity for innovative pedagogical approaches.

## 2. Related Work

In recent years, the integration of intelligent technologies to empower teachers in instructional design has emerged as a prominent research hotspot within the educational technology domain. This section systematically reviews and critiques existing literature across three key dimensions: intelligent instructional design tools, interdisciplinary teaching research, and educational knowledge graphs. The objective is to underscore the distinctiveness and advanced capabilities of the TriQuest platform within this landscape.

**General AI-Assisted Instructional Design Tools.** The advent of large language models has catalyzed the proliferation of numerous AI-powered instructional tools. For instance, tools such as MagicSchool AI and Canva's Magic Write offer extensive pedagogical templates, enabling rapid generation of texts like lesson outlines, activity descriptions, and assessment questionnaires. Concurrently, several studies have investigated the potential of employing general-purpose models, such as ChatGPT, for various instructional design applications (Al-Worafi et al., 2024; Conrad et al., 2024). While these tools demonstrate considerable efficacy in enhancing the efficiency of basic text generation, their inherent limitations are notably apparent. Primarily, their domain-agnostic nature precludes an intrinsic understanding of specific educational frameworks (e.g., China's "Three-Dimensional Goals" and "Core Literacies"), frequently resulting in generated content that deviates from actual pedagogical requirements. Furthermore, their predominant reliance on a "prompt-response" single-turn interaction paradigm proves inadequate for supporting the intricate cognitive processes inherent in instructional design, which necessitate profound critical thinking and iterative refinement. Lastly, an excessive dependence on these tools risks diminishing teacher agency, potentially relegating educators from their role as "designers" to mere "prompt engineers," which ultimately impedes their long-term professional growth and development.

**Intelligent Support for Interdisciplinary Teaching.** In response to the distinctive pedagogical demands of interdisciplinary teaching, a growing body of research has initiated the exploration of more specialized and tailored solutions. For instance, certain investigations have employed artificial intelligence for purposes such as interdisciplinary theme recommendation or the aggregation of relevant resources (Wang Bo et al., 2024). Other research endeavors have concentrated on the systematic analysis and evaluation of extant lesson plans, utilizing natural language processing techniques to discern and extract interdisciplinary components (Yang Xianmin et al., 2024). While these contributions represent advancements in isolated aspects, they are generally characterized by a fragmented approach. These approaches typically address only discrete, singular issues within the broader interdisciplin ary instructional design workflow (e.g., theme identification or resource curation). Consequently, they conspicuously lack a comprehensive, end-to-end solution that seamlessly integrates and links all critical stages, from "topic establishment" and "content integration" to "activity design" and "evaluation and reflection." Moreover, a significant portion of this research remains confined to the prototype validation phase, with a notable paucity of outcomes that offer stable, user-friendly, and fully integrated platforms. This limitation severely hinders their widespread and scalable implementation among frontline educators.

**Construction and Application of Educational Knowledge Graphs.** As a structured knowledge representation technology, Knowledge Graphs (KGs) hold immense potential for applications within the educational sector. Researchers have explored building subject-specific KGs to facilitate knowledge point navigation, recommend learning paths, and enable intelligent Q&A (Lin et al., 2022). Nevertheless, developing KGs tailored for interdisciplinary instructional design presents distinct challenges. 1) Heterogeneous Knowledge Sources: This necessitates the integration of diverse knowledge, including curriculum standards, textbooks, and pedagogical theories, from multiple sources and formats. 2) Interdisciplinary Relationship Mining: The primary challenge is to automatically and accurately identify and represent profound connections between concepts from different disciplines (e.g., shared principles, transferrable methodologies, resonant values). 3) High Construction Costs: Traditional KG construction is labor-intensive due to reliance on extensive manual annotation, resulting in high costs and long development cycles, which hinder adaptation to dynamic curriculum updates. Consequently, the urgent challenge is to leverage new technologies like Large Language Models (LLMs) to construct low-cost, high-quality, and dynamically updatable interdisciplinary educational KGs.

Current solutions exhibit significant limitations in supporting high-quality interdisciplinary instructional design. Generic AI tools often lack specialization, interdisciplinary research prototypes remain fragmented, and traditional pedagogical platforms are deficient in intelligence and interactivity. These limitations underscore the urgent need for a unified, intelligent, and teacher-centric framework for interdisciplinary curriculum design. The TriQuest framework is specifically designed to address this gap. Through a systematic approach, it deeply integrates curriculum standards, domain-specific knowledge, pedagogical processes, and human-computer collaboration, aiming to deliver an effective solution directly applicable in frontline teaching. The homepage of the TriQuest platform is depicted below.

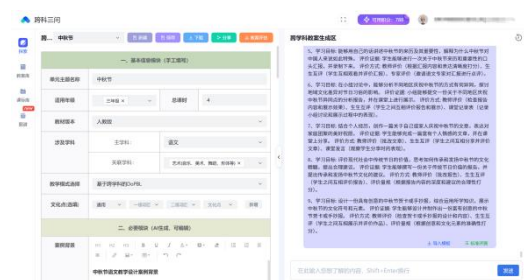

Figure 1: Homepage of the TriQuest.

## 3. Framework

The proposed TriQuest framework offers a unified and scalable approach, designed to empower educators in the efficient creation of high-quality interdisciplinary lesson plans from inception, ensuring alignment with curriculum standards. As illustrated in Figure 2, the TriQuest framework is architecturally delineated into three distinct layers: a data layer, a service layer, and an application layer. This comprehensive framework facilitates the entire pedagogical process—from knowledge input and structured processing to interactive design and evaluative feedback—through a visualized, human-computer collaborative, and intrinsically closed-loop workflow.

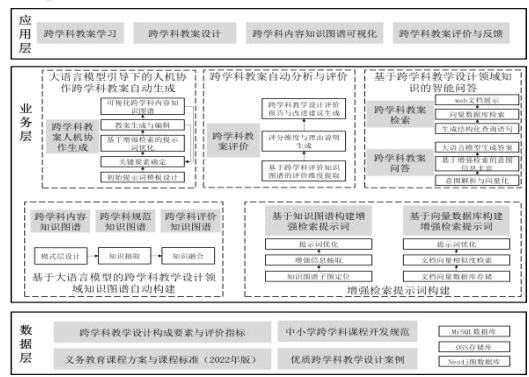

Figure 2: Overview of the TriQuest Framework.

Within the data layer, this research will house the requisite source files for constructing three distinct interdisciplinary teaching design knowledge graphs—specifically, those pertaining to content, pedagogical standards, and evaluation—within an Object Storage Service (OSS). Subsequently, the extracted triplet knowledge will be meticulously stored in a MySQL database before being further imported into a Neo4j graph database for the comprehensive generation of knowledge graphs. Simultaneously, real-time access to both the raw source files and the generated knowledge graphs will be facilitated via open server interfaces, thereby guaranteeing instantaneous data availability and an enhanced interactive experience for users.

At the business layer, the system encompasses five primary modules: interdisciplinary instructional design knowledge graph construction, enhanced retrieval prompt engineering, large language model-guided human-computer collaborative interdisciplinary curriculum design, automated interdisciplinary lesson plan analysis and evaluation, and intelligent Q&A leveraging interdisciplinary instructional design domain knowledge. Specifically, the construction of interdisciplinary content, normative, and evaluative knowledge graphs involves processes such as schema layer design, knowledge triplet extraction, and knowledge triplet fusion, all aimed at optimizing prompt quality and enhancing the accuracy of automated interdisciplinary curriculum design and evaluation. The enhanced retrieval prompt engineering module primarily elevates the quality of interdisciplinary lesson plan retrieval and question-answering by utilizing document vectorization and prompt optimization techniques. The intelligent Q&A model enhances the quality of its responses by interpreting teacher intentions and requirements, then invoking the enhanced retrieval prompt engineering module. The automated interdisciplinary curriculum design module optimizes prompts using the interdisciplinary content knowledge graph, subsequently generating interdisciplinary lesson plans and updated interdisciplinary content knowledge graphs that align with teacher requirements. The automated analysis and evaluation module leverages the interdisciplinary evaluation knowledge graph to call a large language model, which generates scores and justifications for each module of the interdisciplinary lesson plan, ultimately providing actionable recommendations for continuous improvement.

At the application layer, teachers can access and acquire interdisciplinary lesson plan knowledge within the dedicated learning module. The design module facilitates human-computer collaborative generation of interdisciplinary lesson plans, allowing real-time visualization of the corresponding interdisciplinary content knowledge graph. Furthermore, the evaluation and feedback module provides detailed scores, rationales, and suggestions across various dimensions, establishing a clear foundation for iterative refinement.

The proposed system is an AI-powered human-computer interactive interdisciplinary curriculum design system. It features an intelligent lesson plan creation module driven by a large language model (LLM), an LLM-based document question-answering module for an interdisciplinary case library, and an interdisciplinary lesson plan editing module.

**The intelligent lesson plan creation module** facilitates human-computer collaborative design of interactive interdisciplinary lesson plans. It incorporates a built-in prompt template library for various lesson plan components, enabling the generation of specific prompt templates based on interdisciplinary parameters. These templates are refined through both AI-driven and manual optimization, culminating in finalized lesson plan generation prompts. These prompts are then fed into the LLM for question-answering, which subsequently produces the interdisciplinary lesson plan.

**The interdisciplinary case library's LLM document question-answering module** integrates a dynamic and iterative interdisciplinary case repository. This module allows users to search for and view interdisciplinary lesson plans in an online web document reading format. Users can then query the LLM regarding these retrieved lesson plans, receiving LLM-generated answers, which significantly enhances the efficiency of teaching resource acquisition.

**The interdisciplinary lesson plan editing and display module** enables the import of generated interdisciplinary lesson plans into a dedicated editor for manual adjustments. It also supports the construction and interactive visualization of an interdisciplinary knowledge graph, allowing for both display and editing. Finally, the manually reviewed interdisciplinary lesson plans can be downloaded and exported from the lesson plan editor.

### 3.1 Large Model Intelligent Lesson Plan Generation Module

The intelligent lesson plan generation module, powered by a large language model, comprises four sub-modules: prompt template generation, intelligent prompt optimization, lesson plan generation, and lesson plan quality assessment.

The prompt template generation module dynamically constructs structured prompt templates tailored for interdisciplinary lesson plan generation. It achieves this by utilizing user-inputted interdisciplinary curriculum design information in conjunction with a specialized library of prompt templates designed for various lesson plan components. This module encompasses both a comprehensive holistic prompt generation sub-module and a step-by-step prompt generation sub-module. The structured prompt templates are categorized as follows: (i) Case Background Template (ii) Learner Analysis Template (iii) Curriculum Standard Analysis Template (iv) Instructional Content Template (v) Learning Objectives Template (vi) Learning Assessment Design Template (vii) Learning Activities and Design Rationale Template (viii) Theoretical Foundation and Case Design Philosophy Template (ix) Tools and Resources Selection Template.

Furthermore, let the structured prompt templates be defined as C={C1, C2, C3, C4, C5, C6, C7, C8, C9}, where these correspond to {Case Background Template, Learner Analysis Template, Curriculum Standard Analysis Template, Instructional Content Template, Learning Objectives Template, Learning Assessment Design Template, Learning Activities and Design Rationale Template, Theoretical Foundation and Case Design Philosophy Template, Tools and Resources Selection Template}. If user-input interdisci plinary data is denoted by I, then the generation of the prompt template can be formally represented as follows:

$$Di = f(Ci + I)(1 \leq i \leq 4) \quad (1)$$
$$D5 = f(C5 + I + D4) \quad (2)$$
$$D6 = f(C6 + I + D5) \quad (3)$$
$$D7 = f(C7 + I + D5) \quad (4)$$
$$D8 = f(C8 + I + D7) \quad (5)$$
$$D9 = f(C9 + I + D7) \quad (6)$$

Here, $f$ denotes the prompt template generation function. In the context of the step-by-step prompt generation module, template generation processes can proceed independently. Conversely, for the comprehensive holistic prompt generation module, prompt generation will occur sequentially. Specifically, the function $f$ initially monitors changes in the user's template selection. Upon detecting a change, it identifies the relevant template step and parses the prompt embedded within that template. Subsequently, placeholders within the parsed prompt are replaced with their actual values, such as unit theme name or applicable grade level. Finally, upon user interaction with the "generate prompt" button, the processed prompt is converted into a textual format and displayed in the designated input box.

The intelligent prompt optimization module is designed to refine interdisciplinary lesson plan prompts imported into the prompt editing dialog box. This module facilitates both intelligent and manual optimization to ensure the prompts are more precisely aligned with practical requirements. Furthermore, the optimization process can be formally expressed as:

$$Doptimize = g(D) \quad (7)$$

Here, $g$ represents the optimization function, which encompasses two distinct stages: optimization via a large language model and subsequent manual refinement. $Doptimize$ denotes the optimized prompt for lesson plan generation.

Specifically, the function $g$ first defines a set of messages, which includes explicit guiding information for the system role—clarifying the objectives and principles for prompt improvement—and specific questions or requirements for the user role. These messages are then transmitted to the qwen1.5-1.8b-chat large language model. A random seed is set to ensure consistency in the generation process, and a streaming output mode is configured for progressive receipt of generated responses. Furthermore, a streaming output mode is configured to facilitate the progressive reception of generated response content. The function iteratively processes the response stream received from the large model, collecting and concatenating the content from each successful response. In the event of an erroneous response, pertinent error information is duly recorded.

The optimized prompt is then updated in the input box, allowing users to further modify it as needed. Ultimately, the function returns the new, optimized, and concise prompt content.

The intelligent lesson plan generation module is responsible for feeding optimized lesson plan generation prompts into a large language model for interactive question-answering. Subsequently, the large language model produces interdisciplinary lesson plan A, as demonstrated by the following process:

$$A = LLM(Doptimize) \quad (8)$$

More specifically, the function takes a prompt as its input. This prompt is then used to formulate a query that includes precise analytical requirements. For example, the query may direct the large language model to assess students' prior knowledge and skills, general characteristics, and specific learning challenges, and to present these findings in a predetermined format. Ultimately, the system interfaces with the large lan guage model to obtain the generated textual data for the interdisciplinary lesson plan. The lesson plan quality evaluation module meticulously and impartially assesses the generated lesson plan. Leveraging a large language model, it scores the plan against predefined criteria, providing both a score and detailed justifications. These justifications encompass specific evaluations across multiple dimensions, including but not limited to:rationality,comprehensiveness,interdisciplinarity,authenticity,scientific rigor, the challenging nature of activities, organizational effectiveness, the appropriateness of support mechanisms, the holistic nature of outcomes, and overall consistency. These steps enable the lesson plan quality evaluation module to effectively assess the accuracy and applicability of the lesson plan, offer concrete optimization suggestions, and thereby enhance both the quality of the lesson plan and its instructional efficacy.

Building upon this, we define the set of scoring criteria dimensions as S = {s1, s2, ..., s11}, encompassing: rationality, comprehensiveness, interdisciplinarity, authenticity, scientific rigor, activity challenge, organizational effectiveness, appropriateness of support, comprehensiveness of outcomes, overall completeness, and consistency. The aggregate scores and their corresponding justifications are represented by R = {r1, r2, ..., r11}. For each individual scoring dimension, the score and its rationale are computed as follows:

$$ri = \text{LLM}(A, si) \qquad (9)$$

More precisely, upon receiving a lesson plan evaluation request, the function integrates the user-submitted lesson plan content with a predefined scoring rubric template to construct a comprehensive query string. This query string is subsequently transmitted to the large language model for processing and retrieval of the evaluation results. Ultimately, these evaluation results are relayed back to the user. This approach allows the function to conduct a detailed and impartial assessment of the interdisciplinary curriculum design plan against established scoring criteria, furnishing the user with comprehensive scoring justifications.

## 3.2 Interdisciplinary Case Study Document Q&A Module

The large language model's interdisciplinary lesson plan document question-answering module is composed of three main components: a text ingestion module, an interdisciplinary lesson plan retrieval module, and a large language model-based question-answering module. The text ingestion module processes collected interdisciplinary lesson plan files, extracting and storing interdisciplinary knowledge as vector data within a dedicated knowledge base. It subsequently returns a unique file ID, thereby fulfilling the requirements for efficient storage and retrieval of extensive vector datasets.

$$DB = BuildDB(URL) \qquad (10)$$

Here, DB represents the knowledge base, URL denotes the URL of the cross-disciplinary lesson plan file, and BuildDB refers to the text ingestion operation. The BuildDB function specifically utilizes iFlytek's document upload feature, where files or their URLs are submitted via an upload interface to generate a form with file information and essential request headers. A POST request is sent from the system to iFlytek's file upload API endpoint. Upon receiving and processing the request, the server stores the file in the knowledge base. It then segments and vectorizes the document to enhance information retrieval and knowledge extraction, thereby completing the text ingestion process. A file ID is returned by the system after the upload.

Furthermore, the interdisciplinary lesson plan retrieval module queries the interdisciplin ary lesson plan database using the user's input text and specified filter fields. It then retrieves and displays relevant lesson plans in an online web document reading mode. The process for obtaining lesson plan results from the interdisciplinary lesson plan database is as follows:

$$Results = Retrieval(query\_text, filter\_fields) \quad (11)$$

Here, query_text refers to the user's input query, filter_fields represents the set of filtering criteria, and retrieval is the designated retrieval function. Specifically, the function constructs a query object from the preprocessed query text and filter fields. This query object is then used to perform a retrieval within the interdisciplinary lesson plan database to identify matching lesson plans.

The Large Language Model (LLM) Q&A module processes user questions and file IDs, constructing and transmitting WebSocket requests. By invoking the LLM, it delivers highly accurate question-and-answer services. The user's question is defined as an input to the Large Language Model, which then provides answers to the user:

$$answers = LLM(question,\ fileid) \quad (12)$$

Specifically, the LLM function utilizes the iFlytek API to parse user questions and their associated file IDs. It then identifies the necessary documents and content, matching the most pertinent information within the knowledge base to formulate appropriate answers. Should the knowledge base lack relevant information, the function defaults to generating an answer using the large language model as per preset configurations.

### 3.3 Interdisciplinary Lesson Plan Editing and Display Module

The Interdisciplinary Lesson Plan Editing and Display Module consists of an editing and display component and a knowledge graph intelligent construction component. The lesson plan editing and display module imports AI-generated lesson plans into a structured editor for manual review and modification. This process addresses areas where AI may not fully ensure scientific accuracy of specialized knowledge, textual correctness, fluency, coherence, theme matching, content completeness, and punctuation. Finally, the revised lesson plans can be downloaded and exported from the lesson plan editor.

Furthermore, it can be stated as:

$$A_{editor} = Structured\_Editor(A) \quad (13)$$

Here, Structured_Editor refers to the lesson plan structuralization process, and $A_{editor}$ denotes the final set of lesson plans within the structured lesson plan editor. Specifically, the Structured_Editor function utilizes regular expressions to identify key content within the lesson plan. It defines components featuring basic logic and template structures to display and edit this content, including columns such as "Section Name" and "Driving Questions." Furthermore, it offers functionalities for adding, deleting, and resetting table rows.

The intelligent knowledge graph construction module facilitates the creation of interdisciplinary knowledge graphs, visualizing their entities and relationships. Users can further edit edges and nodes (adding, modifying, deleting) and download, export, and save the interdisciplinary knowledge graph. This involves annotating entities in the interdisciplinary lesson plan text within the editor. A large language model then performs semantic analysis on these annotated entities to comprehend their relationships and contextual information. Based on this understanding, the module identifies inter-entity relationships described in the text, extracts knowledge triplets (subject, predicate, object), and ultimately generates the knowledge gr aph. The knowledge graph construction module involves two steps: entity annotation and knowledge triplet extraction:

**Entity Annotation:** To further elaborate, we perform entity annotation on the lesson plan text to derive a comprehensive set of lesson plan text entities. The methodology employed is as follows:

$$A_{annotated} = Clean(A_{editor}) \quad (14)$$

Specifically, the $Clean$ function leverages PaddleNLP, a natural language processing tool, to perform entity annotation. This process employs a named entity recognition model for identification. The resulting lesson plan text entity set comprises two distinct columns: "entity" and "part of speech."

**Knowledge Triplet Extraction**: We conduct semantic analysis on the lesson plan text entities to construct a set of knowledge triplets, denoted as T. Each triplet encompasses a subject, a predicate (attribute), and an object.

$$T = ExtractTriples(A_{annotated}) \quad (15)$$

Specifically, the $ExtractTriples$ function generates a message content for each data batch. This content comprises a CSV string pertinent to the batch and an instructional text. The process mandates the extraction of triplets utilizing dependency parsing principles, followed by invoking the large language model qwen_turbo. The pre-constructed message is then passed to the model.

Subsequently, regular expressions are employed to extract and store the triplets from the large language model's response, thereby forming the knowledge triplet set T.

### 3.4 Model Configuration

To guarantee the framework's flexibility and scalability, the TriQuest platform incorporates a modular backend for both model and resource configuration.

**LLM Interface:** The platform offers flexible integration with various mainstream large language models via APIs, encompassing both commercial models (e.g., iFlytek Spark) and open-source models (e.g., Qwen). This allows different functional modules to dynamically switch between or combine models based on specific requirements for cost, performance, and security.

**Website Deployment:** As the central hub for user interaction within the "Cross-Disciplinary Inquiry" platform, the website deployment module is responsible for seamlessly integrating the platform's front-end and back-end services, making them accessible to users via a dedicated URL. Its core functionalities include: designing and building a user-friendly interface to ensure an optimal user experience; managing business logic, facilitating data interactions, and processing user requests; and ultimately, guaranteeing the platform's stable operation and high-efficiency access.

**Agent Deployment:** Agent deployment is managed through the coze platform, which enables the visual definition of task flows, the construction of comprehensive knowledge bases, and the creation of efficient workflows. This facilitates the automated execution of complex, cross-disciplinary lesson plan generation tasks.

**Backend Services:** The platform's backend is architected around a microservice framework. Key components include: a MySQL database for the persistent storage of user data and lesson plan metadata; a Neo4j graph database dedicated to the storage and efficient querying of knowledge graphs; and an Object Storage Service (OSS) designed for the robust storage of unstructured files, such as lesson plan documents and images.

## 4 Evaluation

To assess the effectiveness of the TriQuest platform within authentic pedagogical contexts, we meticulously designed and execute d a comprehensive user study that integrated both quantitative and qualitative research methodologies.

## 4.1 Experimental Design

Drawing upon the architectural principles and critical technical pathways of intelligent platforms, this research has successfully achieved the following preliminary implementations: **1) Construction and Application of Cross-Disciplinary Content Knowledge Graphs:** This feature empowers teachers to visually inspect the cross-disciplinary knowledge graphs extracted from lesson plans during the design process, providing a clear overview of interconnected concepts. **2) Development and Implementation of Enhanced Retrieval Prompt Generation:** This functionality holds the potential to significantly optimize the effectiveness of teacher questioning by generating more targeted and insightful prompts. **3) Construction and Application of an Intelligent Question-Answering System:** This system enables teachers to learn from exemplary cross-disciplinary lesson plans and provides solutions to challenges encountered during their cross-disciplinary instructional design processes. **4) Human-Computer Collaborative Automatic Generation of Cross-Disciplinary Lesson Plans Guided by Large Language Models:** This capability facilitates human-computer collaboration in cross-disciplinary teaching design, with large language models providing guidance for automated lesson plan generation.

For this study, 43 in-service teachers were recruited to serve as experimental subjects. Prior to the experiment, a five-point Likert scale was administered to assess the participants' previous engagement frequencies in cross-disciplinary teaching design, cross-disciplinary teaching implementation, and the application of intelligent technologies for cross-disciplinary teaching design. The results indicated that the mean frequencies for participants' prior experiences in cross-disciplinary teaching design, cross-disciplinary teaching implementation, and the use of intelligent technology for cross-disciplinary teaching design were 1.86, 1.70, and 1.26, respectively. (These scores are based on a 5-point scale, where 1 denotes "very low" and 2 denotes "low.") These findings collectively suggest that the participants possessed limited prior experience in these areas. The experimental procedure consisted of two main phases: 1) The research team introduced the interface and functionalities of the TriQuest intelligent platform to the participants, who subsequently gained familiarity with the platform through practical, hands-on operation. 2) Participants then self-organized into small groups of four to five members, leveraging the platform's functionalities to engage in various learning activities, including studying interdisciplinary lesson plans, conducting Q&A sessions, and designing their own curricula.

To investigate participants' perceptions of the platform's effectiveness, this study adapted established technology acceptance and behavioral intention scales from Venkatesh et al. (2003) and An et al. (2023) to measure participants' acceptance of and future intention to use the intelligent platform. Concurrently, the study collected participants' reflective journals on their experience of using the platform for human-AI collaborative interdisciplinary lesson design. This qualitative data provided deeper insights and a more comprehensive understanding of their perceptions. In total, 43 valid responses were received for the technology acceptance questionnaire, and 21 valid reflective journals were collected for analysis.

## 4.2 Results and Analysis

**(1) Questionnaire Analysis Results**

As shown in Table 1, the descriptive statis tical analysis yielded the following mean scores: Performance Expectancy (M=3.750), Effort Expectancy (M=3.610), Facilitating Conditions (M=3.343), and Social Influence (M=3.291). The mean scores for both Performance Expectancy and Effort Expectancy exceeded 3.5, indicating a general consensus among participants that the platform was easy to use and effective in supporting their interdisciplinary instructional design efforts. In contrast, the lower mean scores for Facilitating Conditions and Social Influence (below 3.5) suggest that some participants perceived a lack of external support for using the platform and did not feel that key individuals in their environment endorsed its use. This outcome is likely attributable to the study's design, which prioritized investigating the platform's intrinsic effectiveness and thus intentionally minimized intervention during the human-AI collaborative design process, potentially leading to a perceived lack of support. Furthermore, the relatively short duration of platform use and participants' limited experience with it may have prevented the development of habitual use or a supportive social culture, thereby contributing to the lower perceived social influence. Notably, the highest mean score was observed for Behavioral Intention (M=3.833), indicating strong participant agreement and a willingness to continue using the platform for human-AI collaborative interdisciplinary lesson design in the future. In summary, participants generally agreed that the intelligent platform effectively supported their human-AI collaborative interdisciplinary instructional design work and reported a high behavioral intention for its future use.

Table 1: Questionnaire Analysis Results.

| Factor | Maximum | Average | Standard Deviation |
|---|---|---|---|
| Performance Expectancy | 5.00 | 3.75 | 0.937 |
| Effort Expectancy | 5.00 | 3.61 | 0.882 |
| Facilitating Conditions | 5.00 | 3.34 | 0.849 |
| Social Influence | 5.00 | 3.29 | 0.885 |
| Use Intention | 5.00 | 3.83 | 0.781 |

(2) Analysis of Reflective Journals

From the reflective journals, 212 distinct statements were extracted concerning participants' perceptions of the platform's effectiveness for human-AI collaborative interdisciplinary instructional design. These included 98 statements highlighting advantages (see keyword cloud, Figure 2) and 114 statements noting disadvantages or offering suggestions (see keyword cloud, Figure 3). Participants' perceptions of the platform's advantages were categorized into five main themes. The first theme was the enhancement of instructional design efficiency. For educators lacking a strong interdisciplinary background, designing such curricula is often time-consuming and laborious. The platform addresses this by rapidly generating modular lesson structures, providing inspirational ideas and frameworks, and drastically reducing the time and effort required for initial content creation, thereby significantly shortening preparation time. The second theme was overcoming the challenge of topic selection. The platform offered a wealth of thematic suggestions and subsequently generated specific interdisciplinary questions and scenarios, thereby enhancing the richness and interactivity of the resulting lesson designs. The third theme was the provision of a comprehensive, integrated perspective. This was evidenced by the platform's complete set of instructional modules, each containing substantial content—such as accurate student profile analysis, diverse teaching activities aligned with project-based lear ning requirements, and a variety of assessment methods. The fourth theme was its functional richness as a practical tool for daily lesson preparation. Participants highlighted its powerful search functionality, user-friendly interface, strong comprehension capabilities, extensive resources, and its ability to provide teaching materials like charts and diagrams. The fifth and final theme was the facilitation of interdisciplinary knowledge integration. This was demonstrated by its synthesis of multi-disciplinary knowledge to foster fusion and innovation; its breaking down of disciplinary silos, enabling individuals from diverse backgrounds to work across subjects and forge connections between them (e.g., enhancing links between mathematics and other fields); its provision of access to knowledge and insights beyond one's own specialty; and its role in cultivating students' comprehensive application skills. Consequently, a majority of participants concluded that the platform holds significant potential for fostering teacher professional development, encouraging the exploration of novel teaching methods and strategies, improving instructional quality, and thereby enhancing their professional competitiveness.

Participants' identified disadvantages and suggestions for improvement were categorized into four main themes. The first theme was the need for learning objectives to be generated with greater disciplinary specificity. Participants noted that the generated objectives often failed to adequately reflect core disciplinary competencies and required significant revision using appropriate subject-specific terminology. The second theme centered on optimizing the layout and download functionality for lesson plans. Specific issues included disorganized formatting within tables, the requirement for additional plugins to download plans on certain operating systems, and a general need for further user interface refinement. The third theme highlighted the need to enhance the gradation, progression, and flexibility of the suggested interdisciplinary activities. Participants reported that the activity content was sometimes misaligned with student levels—either too advanced or too basic—necessitating significant adjustment. They suggested a greater focus on student cognitive logic and better conceptual sequencing between knowledge points. The activities were also perceived as somewhat formulaic and inflexible, requiring more context-sensitive and practicable designs. The fourth theme concerned limitations in the range of supported textbook versions and educational stages. Notably, the platform lacked alignment with the standard state-compiled textbook series, preventing the generation of relevant activities. Its functionality was also primarily geared towards compulsory education, with limited applicability to high school and other educational levels. Consequently, most participants concluded that targeted modifications to the AI-generated lesson plans were necessary to align with their specific contextual needs and enhance practical applicability.

In summary, participants widely acknowledged that the intelligent platform effectively compensated for their lack of interdisciplinary teaching expertise and enhanced both the efficiency and quality of their interdisciplinary instructional design and teaching practices. Nevertheless, the platform requires further enhancement to better address teachers' needs for personalized interdisciplinary design support. Consequently, future research will focus on implementing continuous improvements targeting these specific areas.

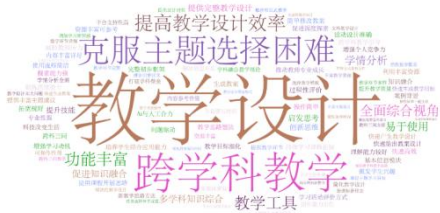

Figure 3: Keyword Cloud for Perceived Advantages of the TriQuest Intelligent Platform.

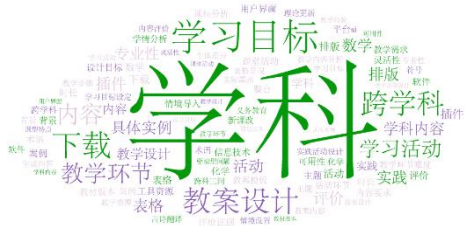

Figure 4: Keyword Cloud for Perceived Disadvantages and Suggestions for the TriQuest Intelligent Platform.

**General Discussion**

Integrating the quantitative and qualitative findings, we conclude that the effectiveness of the TriQuest platform—an AI tool specifically designed for interdisciplinary curriculum design—is strongly supported. Its core mechanism, combining a knowledge graph with a guided workflow, successfully addresses the fundamental challenges teachers face: reluctance to attempt integration (due to apprehension), uncertainty about how to begin (confusion), and lack of methodology to execute it effectively (absence of know-how). The platform functions not only as an efficiency tool but also as a "cognitive scaffold" and an "inspiration catalyst," empowering teachers while simultaneously fostering their professional reflection and growth. The valuable user feedback provides clear direction for the next phase of the platform's iterative development.

## 5 Conclusion and Future Work

**Conclusion**

To address the significant challenges teachers face in interdisciplinary instructional design under the new curriculum reform, this study proposes and implements a unified, scalable intelligent empowerment framework named TriQuest. The framework systematically integrates three core modules: an LLM-based intelligent lesson plan generator, an LLM-powered interdisciplinary case library with document Q&A capabilities, and an interactive lesson plan editor, offering teachers a comprehensive end-to-end solution from thematic exploration to final lesson evaluation. The TriQuest platform demonstrates several key benefits:

Enhancing Lesson Plan Quality and Innovation: By leveraging big data and machine learning capabilities, the platform integrates cutting-edge interdisciplinary research and teaching practices. This provides educators with novel instructional approaches and case studies, thereby enhancing the creativity, contemporary relevance, and overall quality of lesson plans. Optimizing Teaching Resource Integration: Utilizing intelligent analysis and knowledge graph technology, the platform efficiently collects, filters, and organizes relevant interdisciplinary teaching resources. This enables teachers to rapidly identify the most suitable information and cases for their specific needs, significantly reducing the time and effort required for lesson preparation. Increasing Teaching Efficiency and Reach: The automated and intelligent lesson plan generation mechanism allows educators to prepare courses more efficiently, freeing up valuable time for classroom interaction and personalized student guidance. Furthermore, it helps extend the reach of high-quality educational resources to remote and under-resourced areas. Facilitating Implementation of New Curriculum Standards: The platform directly supports the implementation of interdisciplinary thematic learning mandated by the new curriculum standards in primary and secondary education. Adaptive Resource Updates: The sy stem dynamically updates teaching resources, ensuring lesson content remains current by integrating the latest academic research and educational technologies. This promotes continuous improvement in both teaching quality and student learning efficiency.

Our user research results provide strong evidence for the value of the TriQuest framework, demonstrating its capacity to significantly enhance both the efficiency and quality of instructional design while achieving high acceptance and behavioral intention among practicing teachers. This suggests that the design philosophy embodied by TriQuest—transitioning from simply "giving a fish" (direct content generation) to "teaching how to fish" (process-oriented, structured empowerment)—represents an effective pathway for the deep integration of intelligent technology with educational practice. Our work not only presents a viable technical solution to address the practical challenges of interdisciplinary lesson design but also establishes a new design paradigm for the development of future intelligent educational tools.

**Future Work**

While the TriQuest platform has demonstrated initial success, we recognize substantial potential for further development. Our future work will focus on the following key directions:

**Expanding Knowledge System Depth and Breadth:** We plan to continuously enrich the underlying knowledge graph and case library, extending vertical coverage across all educational stages from preschool through high school, while horizontally incorporating more regional textbook versions and emerging disciplines (e.g., AI education, STEAM). Concurrently, we will investigate more advanced techniques for automated knowledge graph updating to maintain alignment with dynamically evolving curriculum content.

**Advancing Human-AI Collaboration Models:** While the current model operates primarily through teacher-led design with AI assistance, we will explore more advanced collaboration paradigms. This includes developing AI capable of personalized adaptation to individual teachers' design habits and styles, and creating an "AI Teaching Assistant" role that can proactively identify potential design issues (e.g., misalignment between activities and learning objectives) and provide constructive feedback, enabling deeper "human-AI pairing."

**Enabling Multimodal Content Generation and Integration:** Recognizing that future instructional design will be inherently multimodal, we will integrate capabilities for generating images, diagrams, animations, and even simple instructional videos into the platform. This will allow educators to create comprehensive multimedia teaching resource packages within a single ecosystem, moving beyond text-based documents.

**Building a Teacher Professional Development Community:** We plan to develop a teacher community integrated with the platform. This community will enable educators to share, critique, and adapt high-quality lesson plans created with the platform, fostering a virtuous ecosystem of "creation-sharing-iteration." Analysis of community data will also allow us to extract collective teaching wisdom and identify innovative patterns, creating a feedback loop for continuous platform enhancement.

**Extending from "Teaching Design" to "Learning Design":** Ultimately, we aim to extend the platform's capabilities to the student side by developing a project-based learning (PBL) navigation tool. This tool would generate personalized learning pathways, inquiry tasks, and resource recommendations for students based on teacher-designed lesson plans, enabling deep integration between teaching and learning and truly embedding interdisciplinary principles throughout the educational process.

We believe that sustained advancement along these directions will enable the TriQuest platform to make significant contributions to educational innovation, teacher empowerment, and the development of students' core competencies.